\documentclass[twocolumn, letter]{emulateapj}
\usepackage{multirow}
\usepackage{graphics}
\usepackage{amsmath, amsthm, amssymb}
\usepackage{epsfig}
\usepackage[hidelinks]{hyperref}
\usepackage{natbib}

\usepackage{graphicx}
\usepackage[dvipsnames]{xcolor}

\begin{document}

\title{The moving groups as the origin of the vertical phase space spiral
 }

\author{Tatiana A. Michtchenko$^{1}$}\email[e-mail: ]{tatiana.michtchenko@iag.usp.br}
\author{Douglas A. Barros$^{2}$}\email[e-mail: ]{douglas.barros@alumni.usp.br}
\author{Angeles P\'erez-Villegas$^{1}$}\email[e-mail: ]{mperez@iag.usp.br}
\author{Jacques R.\,D. L\'epine$^{1}$}\email[e-mail: ]{jacques.lepine@iag.usp.br}

\affiliation{$^1$Universidade de S\~ao Paulo, IAG, Rua do Mat\~ao, 1226, Cidade Universit\'aria, 05508-090 S\~ao Paulo, Brazil\\
$^2$Rua Sessenta e Tr\^es, 125, Olinda, 53090-393 Pernambuco, Brazil}

\date{\today}

\begin{abstract}
Using the \emph{Gaia} data release 2 (DR2), we analyzed the distribution of stars in the close vicinity of the Sun in the full 3D position-velocity space. We have found no evidence of incomplete phase mixing in the vertical direction of the disk, which could be originated by some external events. We show that the vertical phase space spiral $Z$--$V_z$ is produced by the well-known moving groups (MGs), mainly by Coma-Berenices, Pleiades-Hyades and Sirius, when the statistical characteristics (mean, median, or mode) of the azimuthal velocity $V_\varphi$ are used to analyze the distribution in the vertical position-velocity plane. 
This result does not invoke external perturbations and is independent on the internal dynamical mechanisms that originate the MGs. Our conclusions counterbalance current arguments in favor of short-lived (between 300 and 900\,Myr) structures in the solar neighborhood. Contrarily, they support the hypothesis of a longer formation time scale (around a few Gyr) for the MGs.
\end{abstract}


\maketitle

\section{Introduction}
\label{sec:intro}

The \emph{Gaia} data release 2 (DR2) has brought many new results and interpretations for the observed phenomena related to the dynamics of the stars in the Solar neighborhood (SN). One result in particular came as a surprise, the detection by \cite{Antoja_etal2018} of spiral-like structure in the $Z$-direction (perpendicular to the Galactic plane). This structure extends in the ranges of -1\,kpc$< Z <$1\,kpc and -60\,km\,s$^{-1}< V_z <$60\,km\,s$^{-1}$. The authors interpret it as an evidence of incomplete phase mixing in the vertical direction of the Galactic disk, that requires to assume some hypothesis on the origin for this phenomenon. Several works have already proposed explanations, like the influence of a satellite galaxy \cite[]{Antoja_etal2018,2018MNRAS.481.1501B, 2018arXiv180902658B,Laporte+2019}, the buckling instability of the bar \cite[]{2019A&A...622L...6K} and the vertical bending waves in a stellar disk \citep{{2019MNRAS.484.1050D}}.

\cite{Antoja_etal2018} also claim that there is "a strong correlation between the vertical and the in-plane motions of the stars". Since it is generally accepted that the dominant structures in the phase-space distribution of the SN are the moving groups (MGs)
\citep[e.g.][]{antojaEtal2008AA,2018MNRAS.478..228Q}, that statement prompted us to investigate the connection between the vertical structures and the MGs.

We provide a short description of the MGs in the Galactic plane in Sect.\,\ref{sec:sec2}. The MGs have widely been attributed to Galactic resonances, associated with the bar and/or spiral arms, such as the inner and outer Lindblad resonances (ILRs and OLRs) and corotation \citep{perezvillegasEtal2017ApJL,Michtchenko_etal2018b,2018Hattori}.

The distribution of stars in the Galactic plane is compared with the distribution of stars in the $Z$--direction in Sect.\,\ref{sec:sec3}. Our approach is to focus on the arrangement of stars in the phase-space as they are observed, without using any dynamical model to explain the observed distribution. In Sect.\,\ref{sec:sec4}, starting from the known distribution of stars associated with the MGs, we verify that, in a statistical representation of $V_\varphi$ (the velocity in the direction of Galactic rotation) on the $Z$--$V_z$ plane, a spiral-like structure appears more pronounced in the median/mode values than in the mean values. We associate this behavior to the inhomogeneous distribution of the data in the bins with fixed $Z$ and $V_z$ related to the existing MGs. The MGs pull the median/mode towards themselves that produces $V_\varphi$-oscillations between the bins and, consequently, the spiral-like feature on the $Z$--$V_z$ plane.  On contrary to the incomplete vertical phase mixing, which requires a number of assumptions on its origin, our interpretation is based on the well-known fact of existing MGs. There is no need for any external mechanism, but only the known MGs, to explain the observed spiral on the $Z$--$V_z$ plane.

An interesting consequence of our finding concerns the lifetime of the structures in the SN. If we assume that MGs are formed by action of diverse resonances of the spiral arms and/or bar, we can conjecture about the lifetime of these structures. Indeed, the establishment of resonance zones, that are filled with captured stars, is not a fast process. For instance, the azimuthal period of the orbits close to the corotation zone is about 1\,Gyr. Therefore, our result reinforces the idea of longer formation time scales for the structures in the SN, such as the MGs. Consequently, this conclusion not only brings an argument against that the spiral structure in the $Z-V_z$ plane is an evidence of the incomplete phase mixing, which is in favor of short-lived structures (between 300 and 900\,Myr, see \citealt{Antoja_etal2018}),  but it also reinforces the idea of longer formation time scales for the structures in the SN, such as the MGs.


\section{Sample}
\label{sec:sec1}

Our sample consists of stars from the {\it Gaia}  DR2 \citep{Gaia2018A}, which provides positions, parallaxes $\varpi$, proper motions \citep{lindergrenEtal2018AA}, and radial line-of-sight velocities \citep{KatzEtal2018A} for stars with $G<$13\,mag. The sample was restricted to objects with 
parallax errors smaller than $20\%$.  In order to convert the positions, parallaxes, proper motions on the sky, and radial velocities of the selected stars into Cartesian Galactic positions and velocities, we used the \texttt{galpy} python tool \citep{bovy2015ApJS}.
\begin{figure}
	\includegraphics[width=0.99\columnwidth]{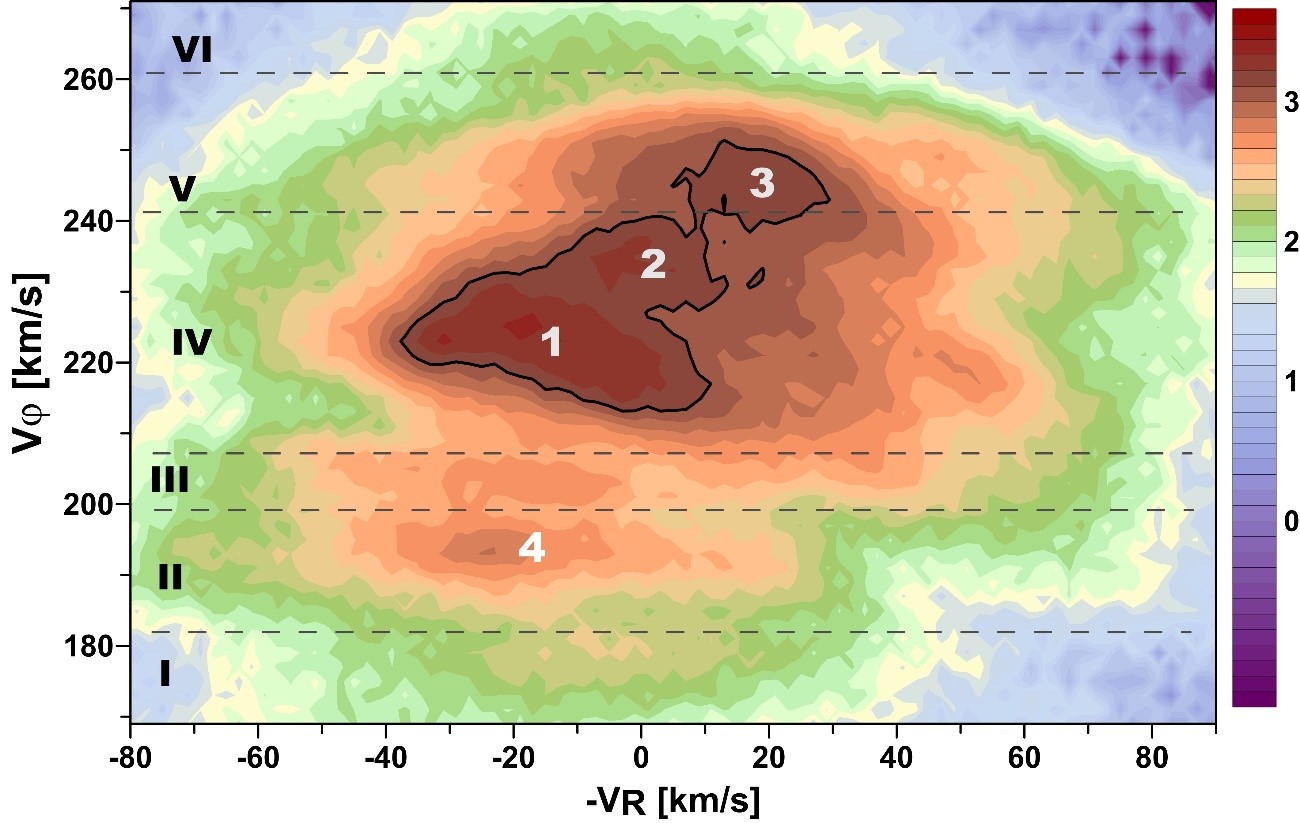}
    \caption{Density of stars in the SN in the $V_R$--$V_\varphi$ plane, in a logarithmic scale, calculated in a grid of 2.0$\times$2.0\,km\,s$^{-1}$. The $V_R$--$V_\varphi$ plane is divided into six zones (see text for details). The location of the main MGs is indicated as: 1 - Pleiades-Hyades, 2 - Coma-Berenices, 3 - Sirius, 4 - Hercules.}
    \label{fig:fig1}

\end{figure}

Following \cite{Antoja_etal2018}, we also transformed the heliocentric rectangular coordinates to the Galactocentric cylindrical coordinates for the stellar positions ($R$,$\varphi$,$Z$) and velocities ($V_R$,$V_\varphi$,$V_z$). We consider that the Sun is placed at the Galactocentric and vertical distances of $R_0=8.0$\,kpc and $Z_0=0.027$\,kpc, respectively, with a circular velocity of $V(R_0)=230$\,km\,s$^{-1}$; these parameters are slightly different from those used in \cite{Antoja_etal2018}. We assume the solar peculiar motion with respect to the Local Standard of Rest as $(11.1,12.24,7.25)$\,km\,s$^{-1}$ \citep{SchonrichBinneyDehnen2010MNRAS}. To compare our results to those presented in \cite{Antoja_etal2018}, we limit the Galactocentric radii to the range of 7.9\,kpc$< R <$8.1\,kpc, the azimuths to $|\varphi|<8^\circ$ from the Sun, and in the vertical position and velocity to $|z|< 0.8$ kpc and $|V_z|<60$ km s$^{-1}$, respectively.
As result, our final catalog contains 838,883 stars.

\section{Sample on the $V_R$--$V_\varphi$ plane and MGs}
\label{sec:sec2}

Figure~\ref{fig:fig1} shows the velocity distribution in the $V_R$--$V_\varphi$ plane. The well-known MGs are easily identified in the figure as streams extended along the horizontal axis, $V_R$. They appear delimited along the $V_\varphi$-axis, producing a ridge-like pattern. These features were analyzed and described in detail in our previous work \citep{Michtchenko_etal2018b}, where we associated them to the action of the corotation and the near-by high-order ILRs and OLRs of the spiral arms. Briefly, 
we summarize below our main results, roughly delimiting in Fig.\,\ref{fig:fig1} the zones of action of these dynamical mechanisms. There are:
\begin{itemize}
\item zone I - stars coming from the inner part of the Galaxy and evolving in the 6/1 ILR;
\item zone II - Hercules stream (\#4 in Fig.\,\ref{fig:fig1}) formed mainly by the stars evolving in the 8/1 ILR;
\item zone III - a weak stream associated with the 12/1 and higher-order ILRs, close to the corotation zone;
\item zone IV - corotation zone harboring the Sun, the MGs of Coma-Berenices(\#2), Pleiades and Hyades (\#1), the main component of the MGs placed at the Local Arm \citep{LepineEtal2017ApJ};
\item zone V - Sirius  group (\#3)  associated to the bulk of overlapping 8/1 and higher-order OLRs, which are stuck to the corotation zone;
\item zone VI - the objects involved in the strong 4/1 OLR.
\end{itemize}

This zone list is one example of interpretation of the possible nature of the MGs. There are several other ones in  the literature; we note that the interpretation is not relevant in the present work. The contour around regions 1, 2 and 3 encompasses partially Pleiades-Hyades, Coma-Berenices, and Sirius, respectively, and represents 90\% of the maximal density of the sample. This contour is used to select stars that clearly belong to each group: Pleiades-Hyades with 114,565 members, Coma-Berenices with 47,722 members, and Sirius with 30,954 members.

\begin{figure*}
	\includegraphics[width=0.9\textwidth]{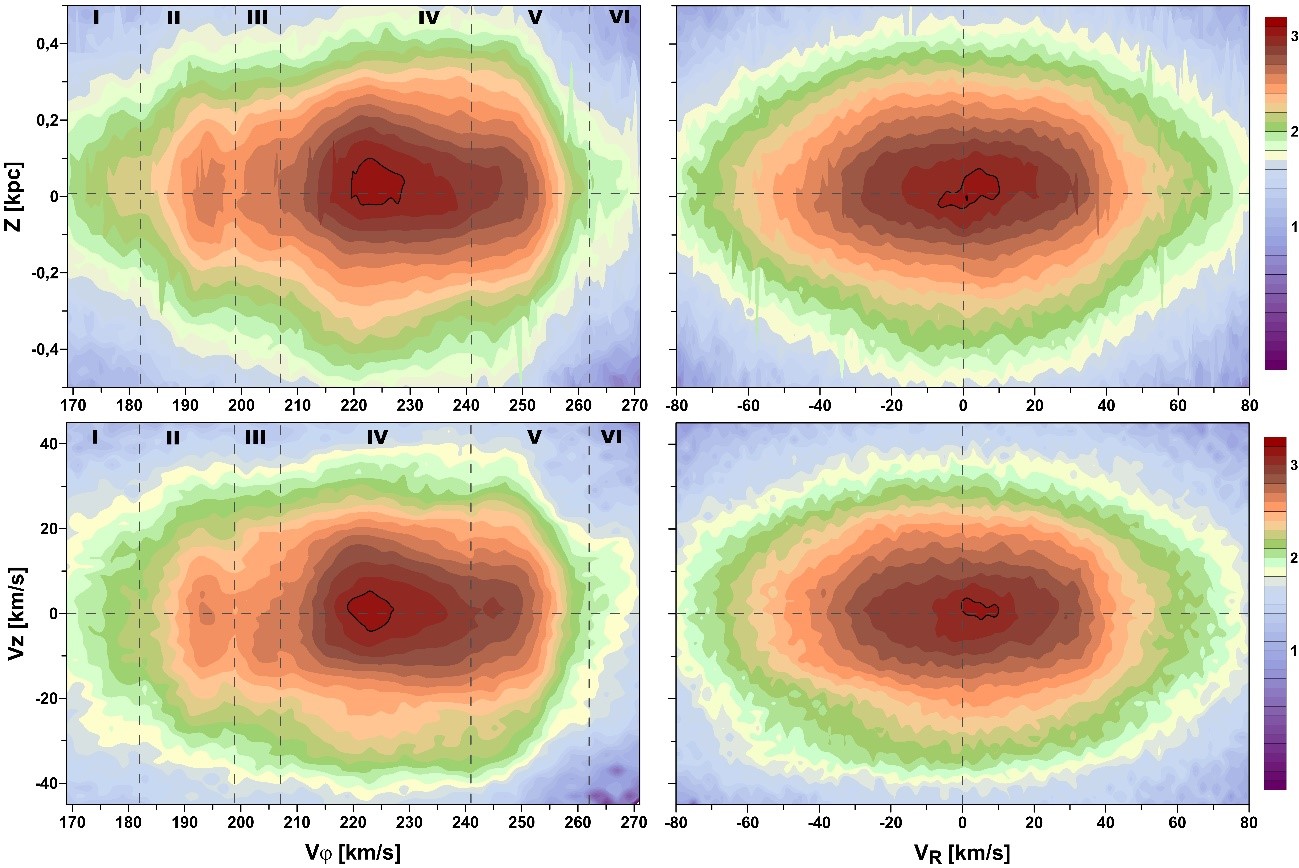}
    \caption{Left column: Density of stars in the SN on the $V_\varphi$--$Z$ (top) and $V_\varphi$--$V_z$ planes (bottom), in the logarithmic scale, calculated on a 2.0\,km\,s$^{-1}\times$0.02\,kpc grid and a 2.0$\times$2.0\,km\,s$^{-1}$ grid, respectively.
    Right column: The same as left panels,  except for $V_R$ along the horizontal axis.
    }
\label{fig:fig2}
\end{figure*}

\section{MGs in the vertical position-velocity direction}
\label{sec:sec3}

Knowing the stellar velocity distribution in the Galactic plane, we now can study the phase and velocity spaces of the MGs outside of the plane. For this, we plot the density of stars in the vertical position-velocity direction as a function of the azimuthal ($V_\varphi$) and radial ($V_R$) velocities in Fig.\,\ref{fig:fig2}. Since the MGs are associated with different ranges of $V_\varphi$,  their respective zones, from I to VI, are easily identified on the planes $V_\varphi$--$Z$ and  $V_\varphi$--$V_z$ (left column). The Hercules stream (zone II) is more evident, since this group is clearly separated from the others. The Hercules members are nearly symmetrically distributed in the vertical direction. The same behavior is observed for the III- and V-zones, which both are stuck to the main component located in the IV-zone. There is no observed correlation between azimuthal and vertical velocities on these density maps (Fig.\,\ref{fig:fig2}), at least, in a large scale (analysis of possible local alterations is out of the scope of this paper).

The vertical position/velocity distribution of the MGs shown in Fig.\,\ref{fig:fig2} (left column) suggests that the motion of stars out of the Galactic plane can be approximated by a one-dimensional oscillator model \citep{binneytremaineGD}.  It is expected that the observed vertical scattering of the groups is correlated with the size of their population, which seems to be correct in the case of the low-density Hercules group.  However, Sirius has approximately only one quarter of the Pleiades-Hyades' population, but the vertical extensions of the corresponding zones IV and V are quite similar. We analyzed the distributions of the heights from the Galactic plane of the stars from Pleiades-Hyades, Coma-Berenices, and Sirius, and we found that the scale-heights of the distributions vary in the narrow range from 127\,pc (Pleiades-Hyades) to 140\,pc (Sirius). This suggests that the vertical excursions of objects can be originated, at least partially, by the Galactic resonances. Indeed, the resonances excite the motion of the group's members and this excitation should be related to the force of the resonance and its chaotic region. The corotation (zone IV) is the strongest resonance, while the chaotic phenomena are stronger in the  Sirius zone (zone V),  where the 8/1 and higher-order OLRs overlap.

Contrarily to what happens along the azimuthal velocity $V_\varphi$, the MGs have no well-defined characteristic values of the radial velocity $V_R$. Indeed, as shown in Fig.\,\ref{fig:fig1}, the MGs are extended along the $V_R$-axis, overlapping in the range between approximately  -20\,km\,s$^{-1}$ and 20\,km\,s$^{-1}$. Due to this fact, the distribution of the stars on the $V_R$--$Z$ and  $V_R$--$V_z$ planes (right column in Fig.\,\ref{fig:fig2}) is more homogeneous and the MGs are not detectable. This difference contributes to the distinct appearances of the azimuthal and radial velocities on the $Z$--$V_z$ plane, as shown in the next section.

\begin{figure*}
	\includegraphics[width=0.95\textwidth]{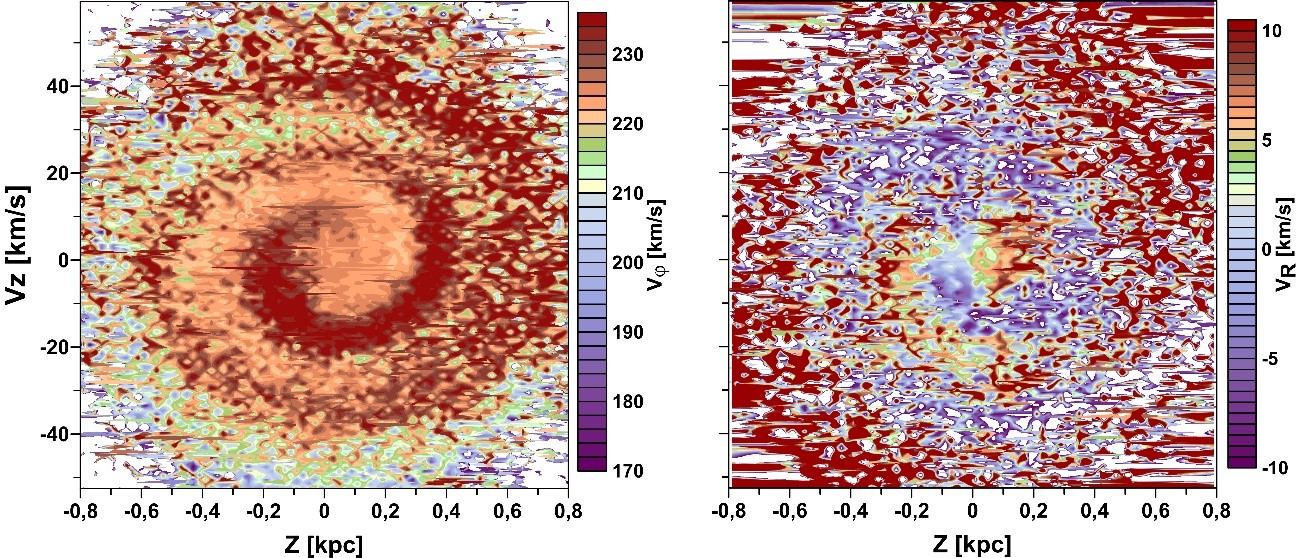}
	\includegraphics[width=0.98\textwidth]{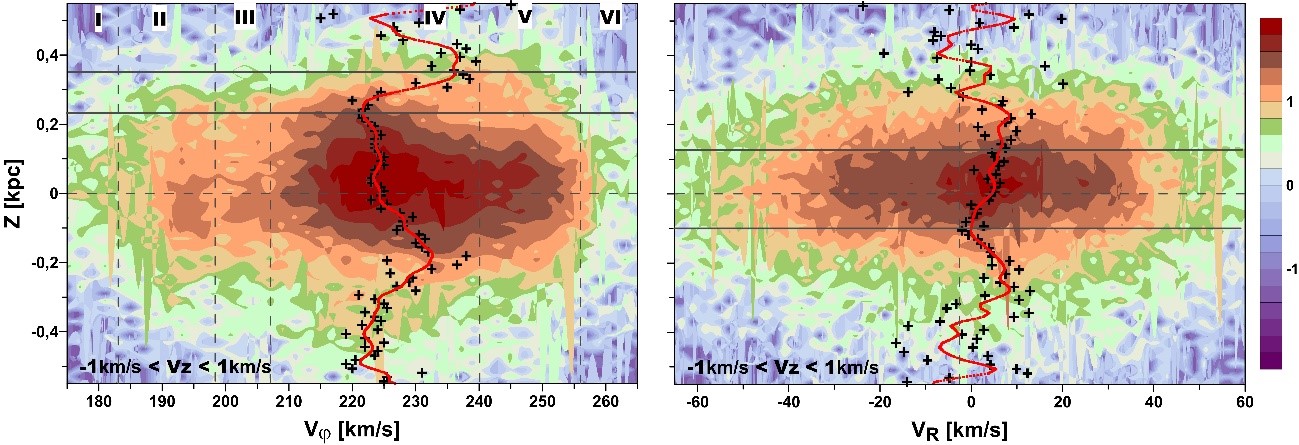}
    \includegraphics[width=0.95\textwidth]{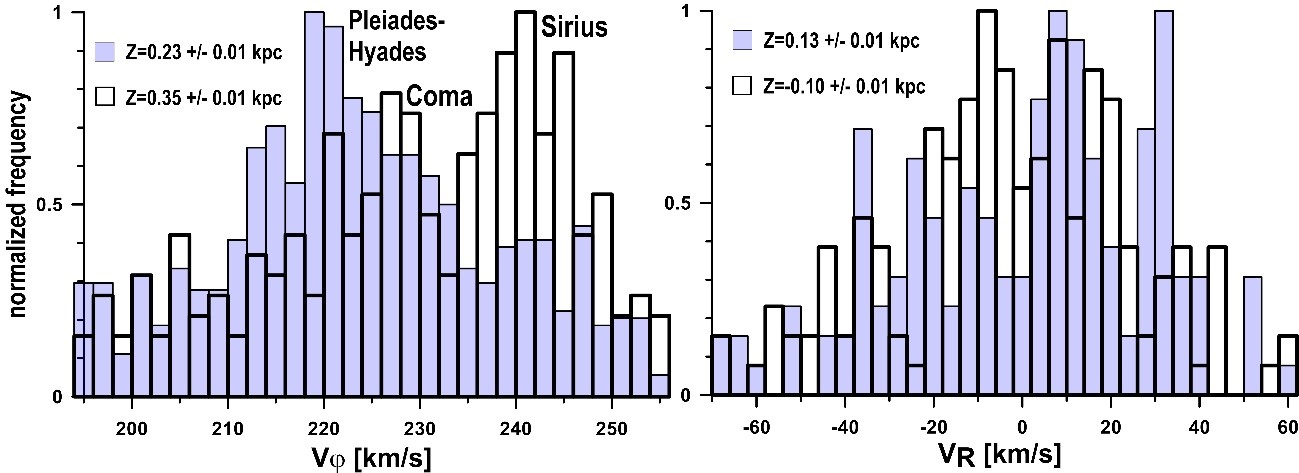}
    \caption{Top row: Mode values of the azimuthal velocity $V_\varphi$ (left) and the radial velocity $V_R$ (right) calculated in bins of 0.02\,kpc$\times$1.0\,km\,s$^{-1}$ on the $Z$--$V_z$ planes.
    Middle row: Same as the top row in Fig.\,\ref{fig:fig3}, considering stars within $-1.0<V_z<1.0$\,km\,s$^{-1}$. The crosses show mode values of the azimuthal velocity $V_\varphi$ (left) and the radial velocity $V_R$ (right) as functions of $Z$. The red curves are the same functions, with  noise filtered by applying a low-pass filter. 
    Bottom row: Normalized frequencies of the stars from the intervals: left - $Z=0.23\pm 0.01$\,kpc (blue) and $Z=0.35\pm 0.01$\,kpc (white); right - $Z=0.13\pm 0.01$\,kpc (blue) and $Z=-0.10\pm 0.01$\,kpc (white). The main MGs are indicated.
    }
\label{fig:fig3}
\end{figure*}

\section{Modes of $V_R$ and $V_\varphi$ in the vertical direction}
\label{sec:sec4}

Figure \ref{fig:fig3} clarifies the mechanisms that contribute to the formation of a spiral on the $Z$--$V_z$ plane, and, at the same time, provides a comparison between the properties of the azimuthal velocity $V_\varphi$ (left column) and the radial velocity $V_R$ (right column). At the top row, the modes of these velocities, obtained in bins of $\Delta Z$=0.02\,kpc and $\Delta V_z$=1.0\,km\,s$^{-1}$, are shown in a linear color scale on the $Z$--$V_z$ plane. These panels reproduce similar information to that one shown by \cite{Antoja_etal2018} in their Fig.\,2, (b) and (c), although they used the medians, not the modes as done here. Several authors prefer to present means and/or medians versions of those figures. We choose to use the modes because this emphasizes the effect of groups present in the sample. A spiral is clearly visible in the $V_\varphi$ representation (top-left panel in Fig.\,\ref{fig:fig3}), while, in the $V_R$ representation (top-right panel), its presence is just marginal, if existent.

The middle row in Fig.\,\ref{fig:fig3} presents the stellar density on the $V_\varphi$--$Z$ \,(left) and $V_R$--$Z$\,(right) planes. These planes are the same of Fig.\,\ref{fig:fig2}\,(top row), except that now we consider the stars confined to a narrow interval of -1\,km\,s$^{-1} <$ $V_z <$ 1\,km\,s$^{-1}$. These planes are useful to analyze the structures located along the $x$--axis on the planes $Z$--$V_z$\,(top row), inside the bins of 0.02\,kpc$\times$2.0\,km\,s$^{-1}$. Using this reduced sample of stars, we calculated the modes of the velocities as functions of $Z$ in bins of 0.02\,kpc, and plotted them on the corresponding planes in Fig.\,\ref{fig:fig3}\,(middle row) by black symbols. Since the mode values are signiﬁcantly spread, we apply a low-pass filter, in order to cut off the numerical noise. The smoothed red curves obtained show clearly an oscillatory behavior, which must produce spiral-like structures when calculated for all velocity values and projected on the $Z$--$V_z$ plane.

Black horizontal lines on the middle panels in Fig.\,\ref{fig:fig3} illustrate the position of bins with fixed $Z$, from which we extract the histograms of objects along  the $V_\varphi$ and $V_R$ directions (bottom row). On the $V_\varphi$--$Z$ plane, we choose two $Z$--values,  which correspond to the minimum value of the $V_\varphi$--mode, at 0.23\,kpc, and the maximum of the $V_\varphi$--mode, at 0.35\,kpc. The $V_\varphi$--mode along the first line (corresponding to the blue-colored histogram on the bottom-left panel) gets its major contribution from Pleiades-Hyades, at $V_\varphi$ around 220 km s$^{-1}$, which is in agreement with Fig.\,\ref{fig:fig1}. The second black line crosses Sirius at $V_\varphi$ around 240 km s$^{-1}$, which gives the major contribution in the white-colored histogram. It also has a minor contribution from Coma-Berenices.

On the middle-right panel in Fig.\,\ref{fig:fig3}, we choose two bins, at $Z$ = -0.10 and $Z$ = +0.13 kpc. We see that the corresponding histograms (bottom-right panel) have several comparable peaks, due to the contribution of the overlapping MGs. As a consequence, the $V_R$--mode does not show well-defined variations when we vary $Z$ (middle-right panel). Thus, the large amplitude mode oscillation of $V_\varphi$ gives rise to the spiral on the $Z$--$V_z$ plane, while the lower amplitude mode oscillation of $V_R$ produces a practically non-existent spiral. 

The discussion above allows us to identify, at least in a first approximation, the MGs that give main contributions to different parts of the vertical phase space spiral-like structure. The white-color histogram in Fig.\,\ref{fig:fig3}\,(bottom-left panel) shows that the Sirius group must be responsible for the outer 'loop' of the spiral-like structure in Fig.\,\ref{fig:fig3}\,(top-left).  This result is supported by the observational evidence that metal-poor stars ([Fe/H]$<$-0.2) are dominating in the external part of the vertical spiral \citep{2018arXiv180902658B}, together with  our interpretation (see Sect.\ref{sec:sec2}), according to which the Sirius group comes from the outer Galactic disk.

Coma-Berenices, located  between Pleiades-Hyades and Sirius, have typical $V_\varphi$-values of around 225--230\,km\,s$^{-1}$ and their main contribution to the spiral-like structure is limited around the origin on the $Z$--$V_z$ plane. The peculiar location of this group, deeply inside the corotation zone, can explain the metal-richness ([Fe/H]$>$0.1) of young stars located in the internal part of the vertical spiral-like structure \citep{2018arXiv180902658B}.

The Pleiades-Hyades group with $V_\varphi$ around 220\,km\,s$^{-1}$ is dominating in population in the SN. We deduce that the valleys corresponding to the lower values of the $V_\varphi$-mode in the spiral-like distribution in Fig.\,\ref{fig:fig3}\,top-left is originated by this group.

\begin{figure*}
    \includegraphics[width=0.95\textwidth]{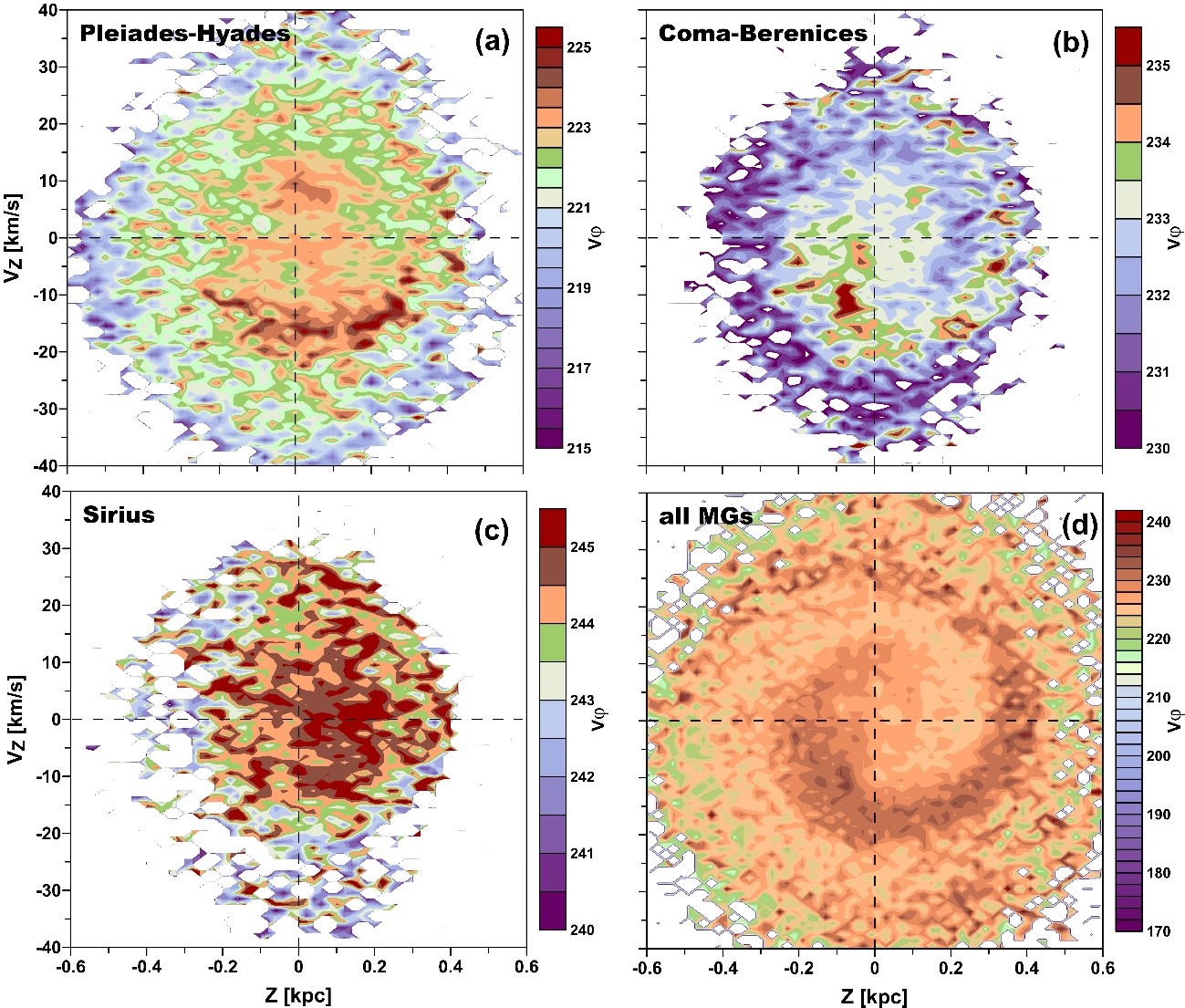}
    \centering

  \caption{Median of $V_\varphi$ of Pleiades-Hyades\,(a), Coma-Berenices\,(b) and Sirius (c), calculated in bins of 0.02\,kpc$\times$1.0\,km\,s$^{-1}$. Each panel is constructed in a linear color scale with stars of only one MG, selected within the contours of regions 1, 2, 3 of Fig.~\ref{fig:fig1}. The panel (d) is the sum of the contributions of the three MGs of the previous panels.
}
\label{fig:fig4}
\end{figure*}

\subsection{A spiral-like structure produced by three MGs}
\label{subsect}

Here we perform a detailed analysis of the groups Pleiades-Hyades, Coma-Berenices, and Sirius, whose members were selected as described in Sect.\ref{sec:sec2}. Our purpose is to show how a spiral feature in the vertical phase space can be produced by the sum of the MGs. In Fig.\,\ref{fig:fig4}, we present the $Z$--$Vz$ plane for each group, with distinct values of the $V_\varphi$--median, given by the color scales beside each panel, from (a) to (c). The panel (d) is the same of the others, obtained by merging the samples of the three groups and then computing the medians in each bin. 

Figure \ref{fig:fig4} shows that the spiral-like feature is only marginally  present in any of the groups on the panels (a)--(c).  The remaining spiral-like features are due to the 'contamination' of one group by the members of others; this effect is vanishing for the Sirius group, which is well-separated from Pleiades-Hyades and Coma-Berenices (see Fig.\,\ref{fig:fig1}).  However, the spiral-like structure appears clearly in the sum of the three samples on the panel (d). This result shows that, to obtain a spiral, nothing else except the well established MGs are needed.

\begin{figure}
	\includegraphics[width=1.0\columnwidth]{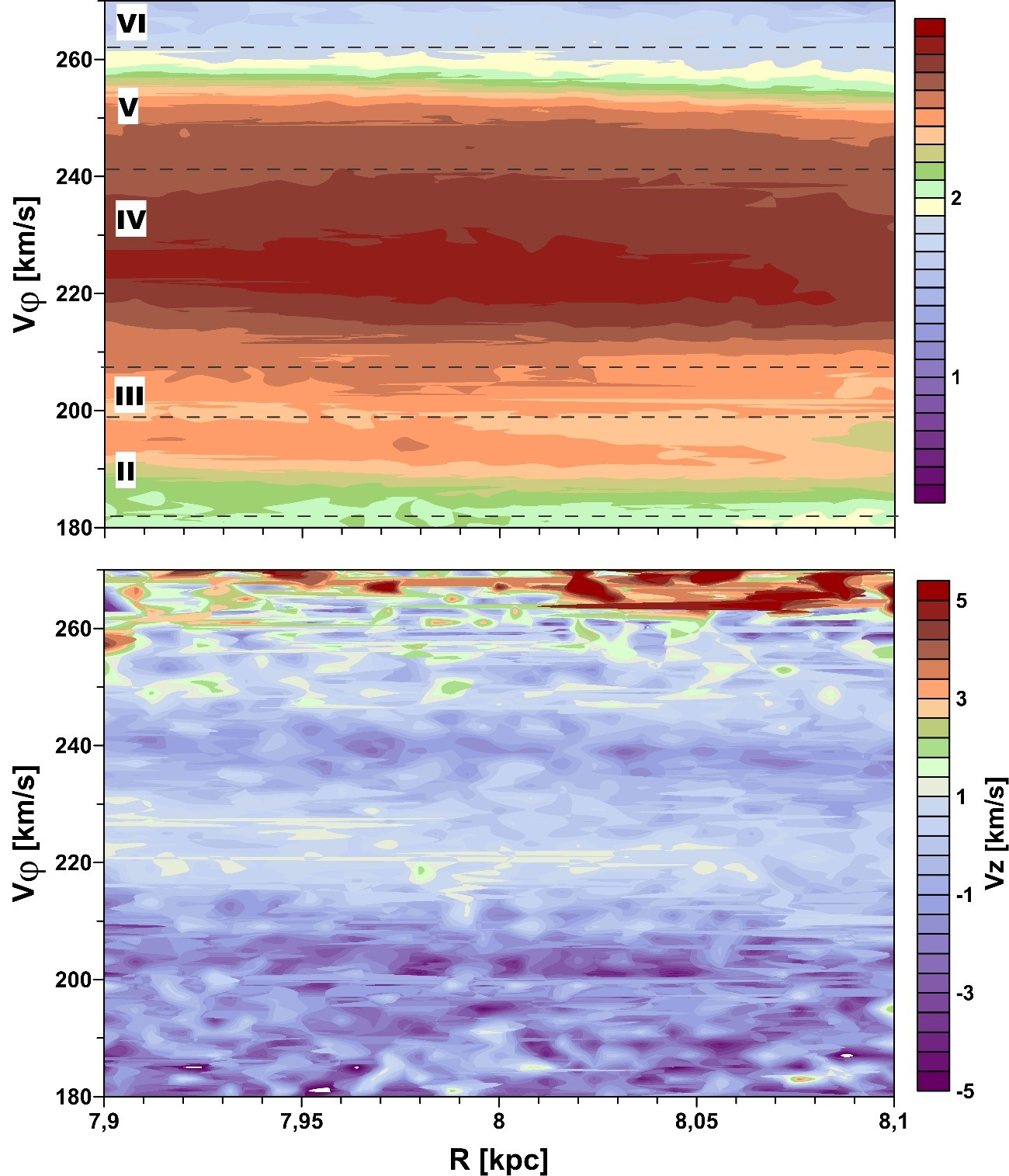}
	
    \caption{Top: Density of the stars from our sample on the $R$--$V_\varphi$ plane, in  logarithmic scale, calculated on a grid 0.004\,kpc$\times$2.0\,km\,s$^{-1}$. The zones of the MGs, from II to VI, are indicated by dashed lines.
    Bottom: Same as on the top panel, except that the median value of $V_z$ is plotted.
   }
    \label{fig:fig5}
\end{figure}

\section{Testing the correlation between the vertical and in-plane motions}
\label{sec:sec6}

The claim made in \citet{Antoja_etal2018}, that the plane and vertical components of the stellar motion are strongly correlated, can be tested on a basis of reciprocity: if the in-plane motion (particularly, the azimuthal velocity $V_\varphi$)  influences the vertical phase-space producing the spiral-like feature (Fig.\,\ref{fig:fig3}\,top-left), the inverse must be true and some structure should appear in the equatorial phase-space also.

We first analyze the distribution of the stars from our sample on the equatorial plane $R$--$V_\varphi$, plotting the star density in bins of  0.004\,kpc$\times$2.0\,km\,s$^{-1}$ as shown in Fig.\,\ref{fig:fig5}\,(top). Even if the $R$-range of the objects from our sample is small (7.9\,kpc$< R <$8.1\,kpc), we can still recognize the ridge-like structure of the SN, already reproduced in  \cite{Antoja_etal2018} (see their Fig.\,2).  The features on the $R$--$V_\varphi$ plane were explained in \citet{Michtchenko_etal2018b} as footprints of the dynamical phenomena, corotation and near-by high-order Lindblad resonances (see Fig.\,2 in that paper). In particular, Hercules, which we associated to the 8/1\,ILR (zone II), is easily distinguished in Fig.\,\ref{fig:fig5}\,top. 

However, the median values of the vertical velocity $V_z$ on the $R$--$V_\varphi$ plane (Fig.\,\ref{fig:fig5}\,bottom) show that there is no a clear  feature, in such a way revealing that, if the coupling between vertical and in-plane motions exists, it should be weak. According to \citet{Khanna2019arXiv}, decreasing sufficiently the $V_z$-range, we will probably find some structures, formed by objects from the very close vicinity to the Galactic plane. These structures would be due to asymmetries, clearly visible in the MG’s distributions in Figs.\,\ref{fig:fig2} and \ref{fig:fig3}. 

\section{Conclusions}
\label{sec:sec7}

We re-examined the distribution of stars around the Sun in the vertical position-velocity phase space. We have observed a spiral-like feature on the $Z$--$V_z$ plane, when the distribution of the mode of the azimuthal stellar velocity, $V_\varphi$, was plotted.  We attributed this behavior to the use of the statistical presentation of $V_\varphi$ (mean, median, or mode) and analyzed its properties in the Galactic plane. The kinematics in this plane shows a highly inhomogeneous distribution for the velocities, characterized by the presence of several massive MGs such as Coma-Berenices, Pleiades-Hyades and Sirius. On the other hand, Hercules has no significant contribution to the vertical phase space spiral due to its low star density. The groups can be roughly separated in bands of specific values of the azimuthal velocity: lower for Pleiades-Hyades and higher for Sirius, with those of Coma-Berenices between them. Plotted out of the Galactic plane, these bands can be still distinguished in vertical positions and velocities.  We show that each of these bands shifts the values of the mean/median/mode in bins in its direction, producing oscillations that originates a spiral-like structure in the $Z$--$V_z$ distribution. Thus, we could explain the 'snail shell' pattern as a kind of behavior produced by the use of the statistical $V_\varphi$-component, which introduces the effects due to the MGs on the $Z$--$V_z$ plane. This result shows that there is no evidence of the incomplete phase mixing in the vertical direction of the Galactic disk and, consequently, excludes the need for engaging external perturbations that would originate it.  Moreover, it counterbalances any argument in favor of short-lived structures in the SN. Indeed, since it needs a few Gyr for stars in the SN to complete a few azimuthal periods, the formation time scale of the MGs should be of the same order. However, this topic should be an issue for a specific study.

\section*{Acknowledgements}

We acknowledge our referee, Dr. James Binney, for the detailed review and for the helpful suggestions, which allowed us to significantly improve the manuscript. This work was supported by the Brazilian CNPq, FAPESP, and CAPES. APV acknowledges FAPESP for the postdoctoral fellowship 2017/15893-1. This work has made use of the facilities of the Laboratory of Astroinformatics (IAG/USP, NAT/Unicsul), funded by FAPESP (grant 2009/54006-4) and INCT-A.




\end{document}